\documentclass[12pt,preprint]{aastex}
\usepackage[update,prepend]{epstopdf}
\usepackage{spr-astr-addons}
\usepackage{lipsum,amsmath,multicol}
\usepackage{amsmath,amssymb}

\RequirePackage{color}

\begin{document}

\title{Anisotropic charged core envelope star}
\slugcomment{}
%% Running heads
\shorttitle{Short article title}
\shortauthors{Autors et al.}

\author{P. Mafa Takisa} 
\affil{Astrophysics and Cosmology Research Unit, School of Mathematics, Statistics and Computer Science, University of KwaZulu-Natal, Private Bag X54001, Durban 4000, South Africa.\\ email: pmafatakisa@gmail.com
}

\author{S.D. Maharaj}
\affil{Astrophysics and Cosmology Research Unit, School of Mathematics, Statistics and Computer Science, University of KwaZulu-Natal, Private Bag X54001, Durban 4000, South Africa.\\ email: maharaj@ukzn.ac.za}

%\author{Subharthi Ray}
%\affil{Astrophysics and Cosmology Research Unit, School of Mathematics, Statistics and Computer Science, University of KwaZulu-Natal, Private Bag X54001, Durban 4000, South Africa.}
%\maketitle

%\email{\emaila}

\begin{abstract}
We study a charged compact object with anisotropic pressures in a core envelope setting. The equation of state is quadratic in the core and linear in the envelope. There is smooth matching between the three regions: the core, envelope and the Reissner-Nordstr\"{o}m exterior. We show that the presence of the electric field affects the masses, radii and compactification factors of stellar objects with values which are in agreement with previous studies. We investigate in particular the effect of electric field on the physical features of the pulsar PSR J1614-2230 in the core envelope model. The gravitational potentials and the matter variables are well behaved within the stellar object. We demonstrate that the radius of the core and the envelope can vary by changing the parameters in the speed of sound.


\end{abstract}

\section{Introduction}
In a general relativistic setting, the study of gravitational behaviour in superdense compact stars is an essential area of research in astrophysics. 
The high central densities in the core provide a physical environment for the relativistic nucleons to convert to hyperons or to generate condensates. 
\cite{Witt1984} and  \cite{Fari1984} studied the first models of the interior of the star with a quark phase.
The real composition for the matter distribution in the core regions remains a question for deeper examination. 
In most studies of relativistic compact stars the core is surrounded by a nuclear crust composed of baryonic matter. 
An example of a such a physical scenario is given by the stellar model of \cite{Sharm2002} with concentric layers of different phases, with inner deconfined quarks and outer less compact baryons. 
In the context of a core envelope model for a relativistic star the matter distributions of the two regions have different physical features. 
In our approach the structure of matter is constrained by equations of state for the envelope and the core. 
In this approach we deal with the macroscopic behaviour of the matter distribution; it is not possible to include the features of the microphysics.
we should point out that in nuclear physics or particle physics, it difficult to specify a single equation of state for the matter distribution that smoothly interfaces the quark matter core with the outer nuclear region. 
Consequently, it is necessary to investigate core envelope models with modification to the equation of state. 

In the past several core envelope models for massive relativistic stars in general relativity have been found. \cite{Sharm2002}, \cite{Paul2005} and \cite{Tike2009}
 considered general relativistic models with different physical features.
 Exact interior solutions with different energy densities in the core and envelope were found by \cite{Durp1969} and \cite{Durp1971}.
  Nonterminating exact series solutions for isothermal neutron star interiors, which are bound and stable were presented by \cite{fulo1988, fulo1989}. 
Models with energy density distribution with parabolic behaviour at the core were generated by \cite{Neg1989, Neg1990}.
\cite{Sharm2001, Sharm2002} showed that their core envelope model could be used to describe the compact X-ray binary pulsar Her X-1: the matter distribution is a quark-diquark mixture in equilibrium. 
Exact solutions to the field equations with the core layer admitting either isotropic or anisotropic pressures were found by \cite{Paul2005}, \cite{Tike2005}, \cite{Thom2005} and \cite{Tike2009}. In these treatments the spacetime geometry was taken to be parabolic, spheroidal or pseudospheroidal.

For a general core envelope model we include the effects of the electromagnetic field, anisotropy and an equation of state. The electric field affects the physical proprieties like mass, radius, density, expansion and gravitational collapse of self gravitating stellar objects. Recently it was observed by \cite{Herr2013} that intense magnetic fields produce anisotropy in neutron stars, white dwarfs and strange stars. Earlier \cite{Rud1972} pointed out that for a realistic stellar model at very high density ranges beyond $10^{14}$ gcm$^{-3}$, the nuclear matter may be anisotropic. A physical requirement is that the matter distribution satisfies a barotropic equation of state throughout the stellar body. In our model the equation of state has a quadratic form in the core region. This ensures that the radial pressures are higher near the stellar centre. Models with a quadratic equation of state have been analysed by \cite{feroze2011}, \cite{maharaj2012} and \cite{Mafa2014b}. The equation of state has a linear form in the envelope. This ensures that the radial pressure in the outer region is lower than the core. Stellar models in general relativity with a linear equation of state have been recently studied by \cite{Mafa2013}, \cite{Mafa2014a} and  \cite{thirukkanesh2013}.

In this paper we present a new core envelope model for astrophysical compact stars by smoothly matching two inner regions each satisfying a different equation of state. The exterior region is characterised by the Reissner-Nordstr\"{o}m metric. We discuss the Einstein field equations in Sect. 2 with charge and anisotropy. In Sect. 3, we  generate the exact solution for the core. The exact solution for the envelope is generated in Sect. 4. In Sect. 5, we present the matching conditions between the two spacetime regions. A detailed physical analysis is undertaken in Sect. 6. Masses and radii for some compact stars are presented in Table \ref{one}. The matter variables are plotted and discussed. We investigate the physical features of the model in connection with the pulsar PSR J1614-2230; the results are presented in Table \ref{two} and Table \ref{three}. 
Results for the gravitational
 redshifts are given in Table \ref{four} and Table \ref{five}. We briefly summarise the results obtained in this paper in Sect. 7.

\section{\label{b1} The model}

For describing the interior spacetime of a stellar charged anisotropic body, the energy momentum tensor must be 
physically relevant. It is given by
\begin{eqnarray}
\label{f1} 
T^{\mu\nu}&=&\text{diag}\left(-\rho-\frac{1}{2}E^2, p_{r}-\frac{1}{2}E^2,\right.\nonumber\\
  & & \left.  p_{t}+\frac{1}{2}E^2, p_{t}+\frac{1}{2}E^2 \right),
\end{eqnarray} 
where the quantities $\rho$, $p_{r}$ and $p_{t}$ and $E^2$ represent the energy density, radial pressure, the tangential pressure and the electric field respectively. These physical quantities describe the matter fields. We also introduce another quantity called the degree of anisotropy $\Delta=p_{t}-p_{r}$. It vanishes for matter with isotropic pressures and  $\Delta=0$. In this study we select a static gravitational field described by the interior metric

\begin{equation}
\label{f2} 
ds^{2} = -e^{2\nu} dt^{2} + e^{2\lambda} dr^{2}
+ r^{2}(d\theta^{2} + \sin^{2}{\theta} d\phi^{2}),
\end{equation}
where $\nu=\nu(r)$ and $\lambda=\lambda(r)$ are arbitrary gravitational functions. The line element $(\ref{f2})$ may be used to model a compact gravitating star such as a superdense star.

We consider a case of a charged gravitating object with anisotropic  pressures in this investigation. 
The Einstein-Maxwell field equations have the form  
\begin{eqnarray}
\label{Queen}
R_{\mu\nu}- \frac{1}{2} R g_{\mu\nu}&=& T_{\mu\nu},\\
\label{Queen}
F_{\mu\nu;\eta}+F_{\nu\eta}{}_{;\mu}+F_{\eta\mu;\nu}&=&0\\
F^{\mu\nu}{}_{;\nu}&=&J^{\nu},
\end{eqnarray}
where $R_{\mu\nu}$ is the Ricci tensor,  $R$ is the Ricci scalar, $T_{\mu\nu}$ is the charged energy momentum tensor, $F_{\mu\nu}$ is the Maxwell tensor and $J^{\nu}$ is the four-current. Hence the effects due to the electric field and pressure anisotropy have been included. Then the Einstein-Maxwell field equations influencing the gravitational interactions in a charged anisotropic star are given by
\begin{eqnarray}
\label{eq:f5} 
 \rho+\frac{1}{2}E^2 &=&\frac{1}{r^{2}} \left[ r(1-e^{-2\lambda}) \right]',\\
 \label{eq:f6} 
 p_r-\frac{1}{2}E^2 &=&- \frac{1}{r^{2}} \left( 1-e^{-2\lambda} \right) + \frac{2\nu'}{r}e^{-2\lambda} ,\\
\label{eq:f7} 
 p_t+\frac{1}{2}E^2 &=&e^{-2\lambda}\left( \nu'' + \nu'^{2}+ \frac{\nu'}{r}\lambda' -\frac{\lambda'}{r} -\nu\right),\\
\label{eq:f8}
 \sigma&=& \frac{1}{r^2}e^{-\lambda}(r^2 E)',
\end{eqnarray}
where primes denote differentiation with respect to the radial coordinate $r$. The quantity $\sigma$ is the proper charge density.
The nonlinear nature of the field equations (\ref{eq:f5})-(\ref{eq:f8}) makes it difficult to integrate them exactly. 
To produce a solution we need to specify the metric functions, prescribe the form of the matter quantities or choose a particular barotropic equation of state.

In core envelope matter configurations the interior of the star is made up of two regions: an inner containing the centre core and an outer envelope region with different pressure profiles.
Therefore to model a core envelope relativistic star we need to divide spacetime into three regions. The three regions comprise of the central core (region $I$, $0\leq r\leq R_{I}$), the neighbouring envelope (region $II$, $R_{I}\leq r\leq R_{II}$) and the exterior of the star (region $III$, $R_{II}\leq r$). The metrics for the three regions have the form 
\begin{eqnarray}
\label{ff2} 
ds^{2}|_{I} &=& -e^{2\nu_{I}} dt^{2} 
 + e^{2\lambda_{I}} dr^{2}\nonumber\\
&&+ r^{2}(d\theta^{2} + \sin^{2}{\theta} d\phi^{2}),\\
\label{ff3} 
ds^{2}|_{II} &=& -e^{2\nu_{II}} dt^{2} + e^{2\lambda_{II}} dr^{2}\nonumber\\
&&+ r^{2}(d\theta^{2} + \sin^{2}{\theta} d\phi^{2}),\\
\label{ff4} 
ds^{2}|_{III} &=& -\left( 1- \frac{2M}{R_{II}} +\frac{Q^2}{R_{II}^2}\right) dt^{2} \nonumber\\
&& + \left( 1- \frac{2M}{R_{II}} +\frac{Q^2}{R_{II}^2}\right)^{-1} dr^{2}\nonumber\\
&& + r^{2}(d\theta^{2} + \sin^{2}{\theta} d\phi^{2}).
\end{eqnarray}
The quantity $Q$ represents the total charge measured by an observer at the infinity.
The metric ($\ref{ff4}$) is the Reissner-Nordstr\"{o}m exterior solution for region $III$ which is exterior to the envelope. The stellar boundary matches smoothly to the Reissner-Nordstr\"{o}m region $III$ which requires vanishing radial pressure at the boundary. The electric field is nonzero at the boundary and matches smoothly with the interior charge.

For physical reasonableness the stellar model should satisfy the following conditions in the core, envelope and exterior: 
\begin{enumerate}
\item[(i)]The gravitational potentials $\nu$ and $\lambda$ and matter variables $\rho$, $p_r$, $p_t$, $E$ and $\sigma$ should be well defined at the centre and regular throughout the star,

\item[(ii)] The energy density $\rho > 0$ $\rho^{\prime} < 0$ in the interior of the star,
\item[(iii)] The radial pressure $p_{r}> 0$, the tangential pressure $p_{t}> 0$, the speed of sound $\frac{dp_{r}}{d\rho}\leq 1$ and the gradient $\frac{dp_{r}}{dr}<0$ in inside the star,
\item[(iv)] At the boundary $p_{r}(R_{II})=0$,
\item[(v)] At the boundary the electric field $E$ must be continuous,
\item[(vi)] The metric functions of the core region $I$ must match with the metric functions of the envelope region $II$, and 
\item[(vii)] The metric functions of the envelope region must match to the Reissner-Nordstr\"{o}m exterior metric.
\end{enumerate}
\section{\label{c1} Region $I$: core}

We use the charged subcase of the exact solution in \cite{maharaj2012}, with a quadratic equation of state, to describe the core region. Including a nonlinear term in the equation of state produces an acceptable model with values for the mass, radius and central density that can be compared with observed stars. Note that a quadratic choice for the equation of state produces higher pressures in the core region. This is the main reason for using the results of \cite{maharaj2012} in region $I$.

In the range $0\leq r\leq { R_{I}}$ we set 
\begin{eqnarray}
\label{SS2}
e^{2\lambda_I}&=&\frac{1+a r^2}{1+b r^2},\\
\label{elect}
E^2&=&\frac{s a^2r^4}{(1+b r^2)^2},\\
\label{SSS2}
p_{r}&=&\gamma\rho^{2},
\end{eqnarray}
where $a$, $b$, $s$ and $\gamma$ are constants. Then the Einstein-Maxwell field equations $(\ref{eq:f5})$-$(\ref{eq:f8})$ give
\begin{equation}
\label{S3}
e^{2\nu_{I}} = B\left(1+a r^2 \right)^{2m_{I}}[1+b r^2]^{2n_{I}}\exp[2F_{I}(r)],\\
\end{equation}
where $B$ is a constant of integration, $a$ and $b$ are parameters related to the central density. The quantity $F_{I}(r)$ is related to the spacetime geometry and has the form
\begin{eqnarray}
F(r) &=& \gamma \left[\frac{2(2b-a)(1+ar^2)+(b-a)}{4(b-a)^{2}(1+ar^2)^{2}}\right]\nonumber\\
&& -s\gamma\left[ \frac{(a-b)^{2}(a r^2+2)-a(2a+s)(1+ar^2)}{4(a-b)(1+ar^2)^{2}}\right]\nonumber\\
&& -s^2\gamma\left[\frac{(a-b)+(3 b)(1+a r^2)}{32(a-b)^{2}(1+a r^2)^{2}} \right]\nonumber\\
&&+\frac{s a r^2(s\gamma -2)}{16b}.\label{SF}
\end{eqnarray}
The form of the equation of state (\ref{SSS2}) yields core densities consistent with earlier treatments; for example compare with the model of \cite{thirukkanesh2013}. The parameters  $\gamma$  and $s$ arise in the exact solutions of \cite{maharaj2012}; we have retained the same parameters for comparison purposes. The various parameters are constrained by the matching across the different regions (see the analysis in Sect. 6).
The constants $m_{I}$ and $n_{I}$ are given by 
\begin{eqnarray}
m_{I} &=& -\frac{s}{8(b-a)}+\gamma  [2(a-b)]^{2}\nonumber\\
&& \times \left(\frac{b^{2}}{(b-a)^{3}}+\frac{b}{(b-a)^{2}}+ \frac{1}{4}\right)\nonumber\\
& & +\frac{s\gamma}{8(a-b)^{3}}\left((a-b)[2s(a-b)+a+b]^2\right.\nonumber\\
&&\left. -6ab +2b^{3}(2a-1) \right),\nonumber\\
n_{I} &=& \frac{(a-b)}{4b}+\gamma  [2(a-b)]^{2} \nonumber\\
&& \times \left(\frac{b^{2}}{(b-a)^{3}}+\frac{b}{(b-a)^{2}}+
\frac{1}{4}\right)+ \frac{sa^{2}}{8b^{2}(b-a)}\nonumber\\
&& +\frac{s\gamma}{16b^{2}(b-a)^{3}}\left[a^{4}(s+4b)+12a^{2}b^{3}-4a^{3}b^2\right].\nonumber\\
\end{eqnarray}
Then the matter variables may be written as
\begin{eqnarray}
\label{S5}
\rho &=& \frac{2(a-b)(3+ar^2)-sa^2r^4}{2(1+ar^2)^{2}},\\
\label{S6}
p_{r} &=& \gamma \rho^{2},\\
\label{S7}
p_{t} &=& p_{r}+\Delta,\\
\label{S8}
\Delta &=& \frac{4r^2(1+br^2)}{1+ax}\nonumber\\
&&\times\left(\frac{m_{I}(m_{I}-1)a^{2}}{(1+ar^2)^{2}}+\frac{2m_{I}n_{I}ab}{(1+ar^2)(1+br^2)}\right.\nonumber\\ 
&&\left. + \frac{2m_{I}a{F}_{I}^{\prime}(r)}{1+ar^2}+\frac{b^{2}n_{I}(n_{I}-1)}{(1+br^2)^{2}} \right.\nonumber\\
& & \left. + \frac{2n_{I}b F_{I}^{\prime}(r)}{1+br^2}+ F_{I}^{\prime\prime}(r)+ F_{I}^{\prime}(r)^{2}\right)
\nonumber\\& &\times \left[\frac{am_{I}}{1+ar^2}+ \frac{bn_{I}}{1+br^2} + F_{I}^{\prime}(r) \right]\nonumber\\
&&+\left[-\frac{2(a-b)r^2}{(1+ar^2)^{2}}+\dfrac{4(1+br^2)}{(1+ar^2)}\right]\nonumber\\
& &+\gamma\left[\frac{2(2b-a)(1+ar^2)+(b-a)}{2(1+ar^2)^{2}}\right]\nonumber\\
&&-\frac{(a-b)}{(1+ar^2)^{2}},\\
\label{S9}
\sigma &=& \frac{4sa^2 r^2(1+b r^2)(2+a r^2)^2}{(1+a r^2)^5}.
\end{eqnarray}
We observe that the gravitational potentials, the matter variables  and the electric field are regular and well behaved in the core region.

\section{\label{c1} Region $II$: the envelope}
We select the charged subcase in the treatment of \cite{Mafa2013}, with a linear equation of state, to describe the envelope region. This class of solutions has been used to describe anisotropic distributions, quark stars and relativistic stars with strange matter.  \cite{Mafa2014a} in a detailed investigation demonstrated that these solutions produce values for the mass and radius consistent with recent estimates for relativistic astronomical objects. This is our main motivation for choosing the exact model of \cite{Mafa2013} for the envelope region $II$.

In the range $R_{I}\leq r\leq { R_{II}}$ we set
\begin{eqnarray}
\label{fS2}
e^{2\lambda_{II}}&=&\frac{1+a r^2}{1+b r^2},\\
\label{elect1}
E^2&=&\frac{sa^2 r^4}{(1+br^2)^2},\\
\label{fSS2}
p_{r}&=&\alpha\rho - \beta,
\end{eqnarray}
where $a$, $b$, $s$, $\alpha$ and $\beta$ are constants. Note that $\alpha$ is related to the speed of sound and $\beta$ is related to the surface density. Then the Einstein-Maxwell field equations $(\ref{eq:f5})$-$(\ref{eq:f8})$ give the metric quantity  
\begin{equation}
\label{S12}
e^{2\nu_{II}} = D\left(1+ar^2 \right)^{2m_{II}}[1+br^2]^{2n_{II}}\exp[2F_{II}(r)],\\
\end{equation}
where $D$ is a constant of integration. The quantity $F_{II}(r)$ is related to the spacetime geometry and has the form
\begin{eqnarray}
F_{II}(r) &=& -\frac{a\beta r^2 +sar^2(1+\alpha)}{8b}.\label{S10}
\end{eqnarray}
The parameters $\alpha$, $\beta$ and $s$ also appear in the exact model of \cite{Mafa2014a} and we retain them for consistency. These parameters become constrained in Sect. 5 by the matching conditions.
The constants $m_{II}$ and $n_{II}$ are given by
\begin{eqnarray}
m_{II} &=& -\frac{(1+\alpha)(s)}{8(b-a)}+\frac{\alpha}{2}\nonumber\\
n_{II} &=& \frac{(1+\alpha)(a-b)}{4b}+\frac{\alpha}{2}
+ \frac{\beta (a-b)}{4b^{2}}\nonumber\\
&&  + \frac{sa^{2}(1+\alpha)}{8b^{2}(b-a)}.
\end{eqnarray}
Then the matter variables become
\begin{eqnarray}
\label{S13}
\rho &=& \dfrac{2(a-b)(3+ar^2)-sa^2r^4}{2(1+ar^2)^{2}},\\
\label{S14}
p_{r} &=& \alpha \rho-\beta ,\\
\label{S15}
p_{t} &=& p_{r}+\Delta,\\
\label{S16}
\Delta &=& \frac{4r^2(1+br^2)}{1+ax}\nonumber\\
&&\times\left[\frac{m_{II}(m_{II}-1)a^{2}}{(1+ar^2)^{2}}+\frac{2m_{II}n_{II}ab}{(1+ar^2)(1+br^2)}\right.\nonumber\\ 
&&\left. + \frac{2m_{II}a{F}_{II}^{\prime}(r)}{1+ar^2}+\frac{b^{2}n_{II}(n_{II}-1)}{(1+br^2)^{2}} \right.\nonumber\\
& & \left. + \frac{2n_{II}b F_{II}^{\prime}(r)}{1+br^2}+ F_{II}^{\prime\prime}(r)+ F_{II}^{\prime}(r)^{2}\right]
\nonumber\\& &\times \left[\frac{am_{II}}{1+ar^2}+ \frac{bn_{II}}{1+br^2} + F_{II}^{\prime}(r) \right]\nonumber\\
&&+\left[-\frac{2(a-b)r^2}{(1+ar^2)^{2}}+\dfrac{4(1+br^2)}{(1+ar^2)}\right]\nonumber\\
& & -\frac{1}{(1+ar^2)^{2}}[(a-b)-\beta(1+ar^2)^{2}]\nonumber\\
& & -\alpha\left[\frac{((a-b))(3+ar^2)}{(1+ar^2)^{2}}\right],\\
\label{accia1}
\sigma &=& \frac{4sa^2 r^2(1+b r^2)(2+a r^2)^2}{(1+a r^2)^5}.
\end{eqnarray}
The gravitational potentials, the matter variables and the electric field are continuous and regular in the envelope.
\section{\label{d} Matching conditions}
The line element ($\ref{ff2}$) for the core and ($\ref{ff3}$) for the envelope must match at $r=R_{I}$. This generates the conditions

\begin{eqnarray}
\label{S17}
e^{2\lambda_{I}({R_{I}})}&=&e^{2\lambda_{II}({R_{I}})},\\
\label{S18}
e^{2\nu_{I}({ R_{I}})}&=&e^{2\nu_{II}({R_{I}})}.
\end{eqnarray}
The line elements ($\ref{ff3}$) and ($\ref{ff4}$) should match smoothly at $r=R_{II}$. This produces the conditions

\begin{eqnarray}
\label{S19}
e^{2\lambda_{II}(R_{II})}&=& \left(1- \frac{2M}{R_{II}} +\frac{Q^2}{R_{II}^2}\right)^{-1},\\
\label{S20}
e^{2\nu_{II}(R_{II})}&=& 1- \frac{2M}{R_{II}} +\frac{Q^2}{R_{II}^2}.
\end{eqnarray}

The radial pressure $p_r$ has to be continuous  at $r=R_{I}$ giving
\begin{eqnarray}
\label{S21}
\gamma \rho^{2}({ R_{I}})&=&\alpha \rho({R_{I}}) - \beta.
\end{eqnarray}
The radial pressure should vanish at the surface of star $r=R_{II}$ giving
\begin{eqnarray}
\label{S22}
\alpha \rho(R_{II}) - \beta &=& 0,
\end{eqnarray}
Smooth matching of the electric $E$  across the stellar boundary gives
\begin{eqnarray}
\label{S23}
Q^2 &=&R_{II}^4 E^2.
\end{eqnarray}
In the above $a$, $b$, $B$, $D$, $R_{I}$, $R_{II}$, $M$, $s$ , $Q$, $E$, $\alpha$, $\beta$ and $\gamma$ are parameters. 
Note that the equations ($\ref{S17}$)-($\ref{S23}$) comprise an undetermined system of seven equations in thirteen unknowns. We may write particular parameters in terms of others. 

The total charge $Q$ in the star is given by ($\ref{S23}$). The other physically relevant quantities are the mass ($M$) and the radius ($R_{II}$) of a compact object. From the system ($\ref{S17}$)-($\ref{S22}$), we find that

\begin{eqnarray}
\label{mer1}
  M &=& \frac{((a-b)(1+a R_{II}^{2})+2a^2 s )R_{II}^{3}}{2( 1+a R_{II}^2)^2},
\end{eqnarray}
which is the total mass of the core and the envelope. In addition we find the quantity 

\begin{eqnarray}
\label{S25}
R_{II}&=&\left(\frac{\alpha(a-b)-2\beta}{a(s\alpha+2 \beta)}\right.\nonumber\\
&& + \left.\frac{(\alpha^2(a-b)(a-b+6s)}{a(s\alpha+2 \beta)}\right.\nonumber\\
&& + \left. \frac{2\alpha(4a-4b-s)\beta)^{1/2}}{a(s\alpha+2\beta)}\right)^{1/2}
\end{eqnarray}
which is the radius of the star. Then the constants $b$, $D$ and $B$ can be written in terms of $M$ and $R_{II}$. These constants are given by

\begin{eqnarray}
\label{S24}
b&=& - \frac{\alpha (1+a R_I^2)^2}{2\gamma(3+a R_I^2)}\nonumber\\
& & + \frac{a\gamma(6+a R_I^2(2-s R_I^2))}{2\gamma(3+a R_I^2)}\nonumber\\
&& -\frac{(1+a R_I^2)^2\sqrt{\alpha^2-4\beta\gamma}}{2\gamma(3+a R_{I}^2)}\\
\label{S26} 
D&=&\frac{(R_{II}-2 M+R_{II}^3 E^2) \left(1+ a R_{II}^2 \right)^{-\alpha /2} }{R_{II}}\nonumber\\
&& \times \left(1+ b R_{II}^2 \right)^{-\frac{\beta (b-a)}{4 b^2 c}-\frac{a (\alpha +1)}{4 b}+\frac{3 \alpha }{4}+\frac{1}{4}}\nonumber\\
&&\times \left(1+ a R_{II}^2\right)^{\frac{s a(-1+\alpha)}{8(a-b)}}\nonumber\\
&&\times e^{-\frac{a \beta R_{II}^2}{4 b c}} \left(1+ b R_{II}^2\right)^{\frac{s a^2(1+\alpha)}{8(a-b)b}}\\
\label{S27}
B&=&\frac{D\exp \left(-\frac{\gamma (a (4b R_{I} -2 a R_{I}-3)+5 b)}{2 (a
   R_{I}+1)^2}-\frac{a \beta {R_{I}}^2}{2 b}\right)}{ \sqrt[4]{a {R_I}^2-{R_I}^2 (a-b)+1}}\nonumber\\
	&&\times\exp\left(\frac{s a\gamma(-3a+b+2a(-2a+b)R_I^2)\times}{16(a-b)^2(1+a {R_I}^2)^2}\right.\nonumber\\
	& & \left. (-4b+a(4+s))\right)\nonumber\\
&&\times \left(1+ b{R_I}^2 \right)^{\frac{ \gamma (a-3b)^2}{4 (a-b)}+\frac{a (\alpha +1)}{2
   b}-\frac{\beta (a-b)}{2 b^2}-\frac{a}{4b}+\frac{3 \alpha }{2}}\nonumber\\
&&\times\left(1+ a R_{I}^2\right)^{\alpha -\frac{ \gamma (a-3b)^2}{4(a-b)}}\nonumber\\
&&\times \left(1+ a{R_{I}}^2 \right)^{\frac{2s\alpha a(a-b)^2}{8(a-b)^3}}\nonumber\\
&&\times \left(1+ a{R_{I}}^2 \right)^{\frac{(-4s\gamma a^2 b(a-3b)(a-b)+s^2a^4)^2}{8(a-b)^3b^2}}\nonumber\\
&&\times \left(1+ b{R_{I}}^2 \right)^{-\frac{2s a^2b(a-b)^2}{8(a-b)^3}}\nonumber\\
&&\times \left(1+ b{R_{I}}^2 \right)^{-\frac{(-4 s\gamma a^2b(a-3b)(a-b)+s^2a^4)^2}{8(a-b)^3b^2}}.
\end{eqnarray}
In ($\ref{S23}$)-($\ref{S27}$) the constants $a$, $\alpha$, $\beta$, $\gamma$, $s$, $E$ and $R_{I}$ are free parameters.

The matching conditions found in this section apply to a charged star with a Reissner-Nordstr\"{o}m exterior. 
When $s=0$ then the electric field vanishes and the star is uncharged with a Schwarschild exterior. If we set $s=0$ in ($\ref{S17}$)-($\ref{S23}$) then we regain the matching conditions established by \cite{takisa2016} for a hybrid star with uncharged core and envelope.

\section{\label{e} Physical analysis}
It is possible to demonstrate that the core envelope model found in this treatment is consistent with astronomical observations. To this end we generate masses for particular stars. We introduce the following new parameters
$\tilde{a}=a{\cal T}^{2}$, $\tilde{b}=b{\cal T}^{2}$, $\tilde{s}=s{\cal T}^{2}$, which scale $a$, $b$ and $s$. In the above
${\cal T}$ is a parameter with dimension of $length$, which assists in comparing with earlier results. 
With the choice of $\tilde{a}=1$, $\beta=0.00162$, $\tilde{s}=14.5$ and the values of $\tilde{b}$, $\alpha$, $\gamma$ given in Table 1 we can generate specific numerical quantities for the core radius $R_{I}$, the envelope radius $R_{II}$ and the stellar mass $M$ for the objects PSR J1614-2230, Vela X-1, PSR J1903+0327, Cen X-3 and SMC X-1. 
We observe that the values for the stellar radius $R_{II}$ is in the range $8.25- 9.21$ km, and values for the mass is in the range  $1.40 - 2.13$ $M_{\odot}$. We emphasize the consistency of these range of values with the treatment of \cite{Mafa2014a} who studied exact solutions to the field equations with equation of state in the absence of charge. We note that similar value for the mass were obtained by \cite{gangopadhyay2013}, \cite{Mafa2014a} and \cite{Mafa2014b}. We have also provided the mass-radius $\frac{M}{R_{II}}$ relationship for five compact astronomical objects in Table \ref{table1-regular}. Observe that the \cite{Buch1959} limit $\frac{2M}{R_{II}}< \frac{8}{9}$ is satisfied in all cases.

\begin{figure}[!ht]
    \centering
		\vspace{0.1cm}
        \includegraphics[width=8.1cm,height=6.1cm]{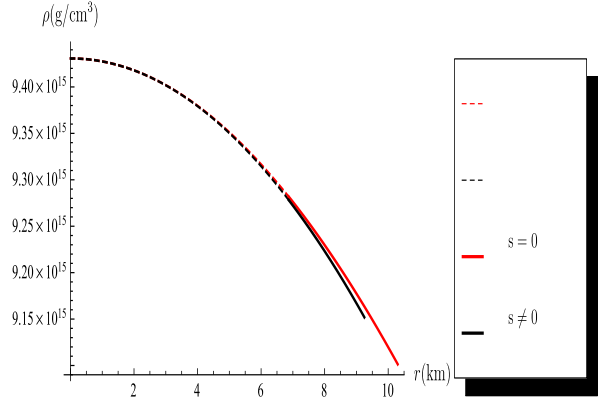} \caption{Energy density}
				\label{one}
\end{figure}

\begin{figure}[!ht]
\centering
 \includegraphics[width=7.2cm,height=4.4cm]{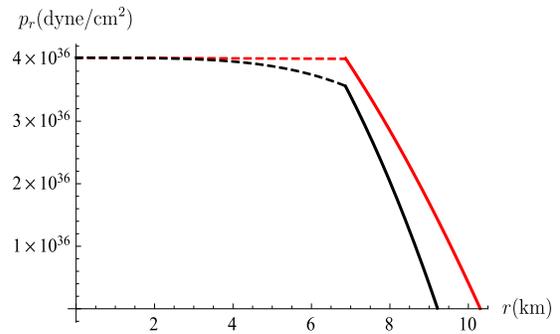}  \caption{Radial pressure $p_{r}$ versus radius $r$.}
 \label{two}
\end{figure}

\begin{figure}[!ht]
\centering
 \includegraphics[width=8cm,height=5cm]{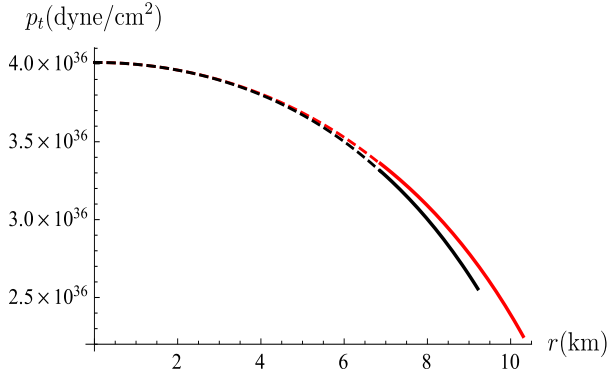}  \caption{Tangential pressure $p_{t}$ versus radius $r$. }
 \label{three}
\end{figure}

\begin{figure}[!ht]
\centering
 \includegraphics[width=7.5cm,height=4.5cm]{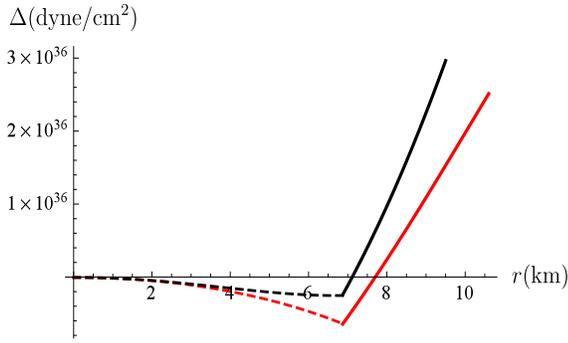}  \caption{Anisotropy $\Delta$ versus radius $r$.}
 \label{four}
\end{figure}

\begin{figure}[!ht]
\centering
 \includegraphics[width=8.2cm,height=5.2cm]{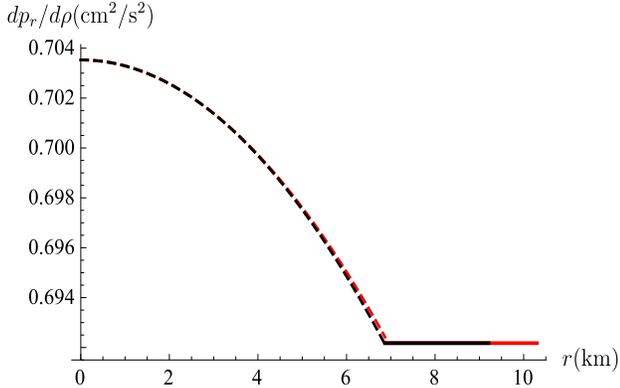}  \caption{Speed of sound $\frac{dp_r}{d\rho}$ versus radius $r$.}
 \label{five}
\end{figure}

\begin{figure}[h!]
\centering
 \includegraphics[width=7.5cm,height=4.5cm]{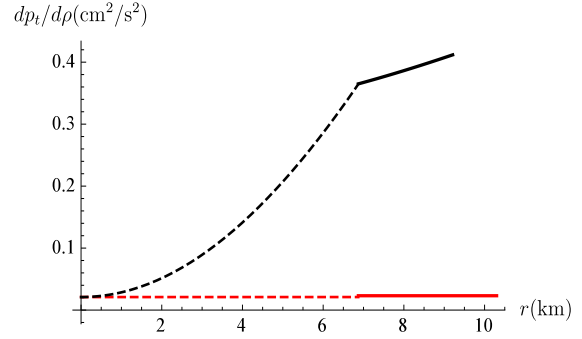}  \caption{The quantity $\frac{dp_t}{d\rho}$ versus radius $r$.}
 \label{six}
\end{figure}

\begin{figure}[h!]
\centering
 \includegraphics[width=7.2cm,height=4.2cm]{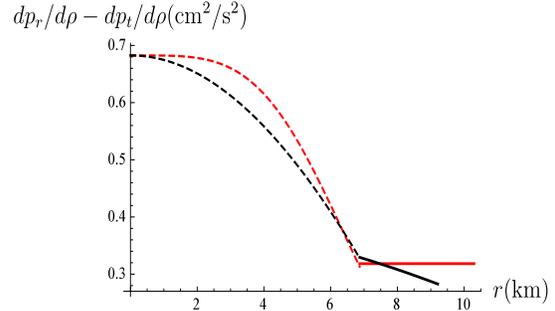}  \caption{ The quantity $(\frac{dp_r}{d\rho} -\frac{dp_t}{d\rho})$ versus radius $r$.}
 \label{seven}
\end{figure}

\begin{figure}[h!]
\centering
 \includegraphics[width=8.2cm,height=5.2cm]{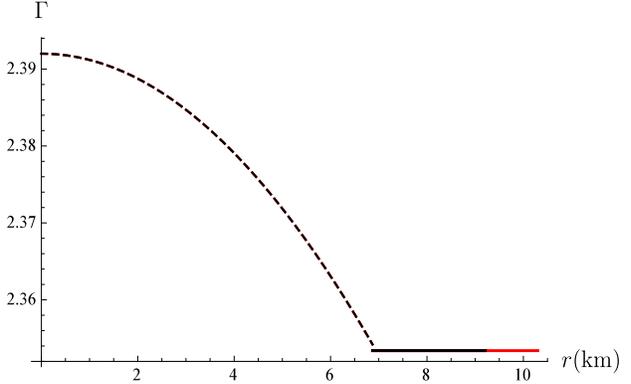}  \caption{Adiabatic index $\Gamma$ versus radius $r$.}
 \label{eight}
\end{figure}

\cite{gangopadhyay2013} studied the astronomical object PSR J1614-2230, which has a feature that the accurate measurement of its mass yields the strongest constraint on the equation of state of superdense star. 
The compactification factor for the object PSR J1614-2230 is $\frac{M}{R_{II}}=0.191$ for the uncharged case and $\frac{M}{R_{II}}=0.231$ for the charged case. For both cases these values lie in the range for neutron stars, compact objects and ultra compact stellar bodies. Comparable values of the compactification factors have been found in the treatment of \cite{Mafa2014a} using a linear equation of state, and by \cite{Mafa2014b} while investigating the field equations with the nonlinear equation of state. 
We now consider the physical analysis of the object PSR J1614-2230 in a core envelope setting with charged matter when $s\neq 0$. In order to study various physical conditions throughout the stellar body we take mass to be 2.13 $M_{\odot}$ and the radius of the envelope to be $R_{II}=9.21 $ km. Similar radial values were found in previous treatment of \cite{Mafa2014a}, \cite{Mafa2014b} and \cite{azam2015}, who studied the stability of the \cite{Mafa2014a} model. We have chosen the core radius to be the two-thirds of the envelope $R_{I}=\frac{2}{3}R_{II}=6.87$ km. 
Table 1 represents values for the charge and uncharged matter. The first set of values represents the uncharged case, and the bracketed values are the corresponding values for charged matter when $s\neq 0$. The variation of speed of the sound $dp_{r}/d\rho$ in Table 1 for different values of $b$, can be viewed as the response of the spacetime geometry to a variation of equation of state for given values of mass and radius of a compact star. The choice of parameters in the model fixes the values of the core radius $R_I$ and the envelope radius $R_{II}$. 
In Table 2 we retain the value of stellar mass to be 2.13 $M_{\odot}$ and the radius of the envelope to be $R_{II}=9.21 $ km.
It shows that the core radius varies and the envelope radius is fixed. In presence of charge the core becomes smaller and more compact. In Table 3 we use the stellar mass value of 2.13 $M_{\odot}$ and the radius of the core is chosen to be $R_I=6.87$ km.
We allow the parameters $\alpha$ and $\gamma$ to vary and the same core radius, in order to keep the value of the mass 2.13 $M_{\odot}$ constant in Table 3. 
We found that when $\alpha$ increases, the same envelope radius $R_{II}$ increases as well so that the envelope radius is changing. Therefore the envelope becomes larger and less compact. 
We have also included Tables 4 and 5 for gravitational redshifts for the region $I$ and region $II$ for the stellar object PRSJ1614-2230. We have included redshift values for both uncharged and charged models. The redshift values for the region $I$ are larger than region $II$. The values are physically reasonable and consistent with other investigations.

The potentials $e^{2\lambda}$ and $e^{2\nu}$ are regular within the compact object with smooth matching between the core, envelope and the Reissner-Nordstr\"{o}m exterior. 
We represent in Figures 1-8 the profiles of the energy density $\rho$, the radial pressure $p_r$, the tangential pressure $p_t$ the measure of anisotropy $\Delta$, the speed of sound $\frac{dp_r}{d\rho}$, the quantity $\frac{dp_t}{d\rho}$, the quantity
$\frac{dp_r}{d\rho}-\frac{dp_t}{d\rho}$ and the adiabatic index $\Gamma$. 
These figures have been plotted for the particular parameter values $\tilde{a}=1$, $\alpha=0.1021$, ${R_I}=6.87$, $\gamma=0.0781$, $\beta=0.00162$ and $\tilde{s}=14.5$. In Figures 1-8 the red dots and continuous line represents the uncharged matter profile and the black dots and continuous line represents the charge matter profile. 
When $R_{I}=R_{II}$, all figures profiles have a smooth matching between the core and the envelope. 
All the matter variables are  regular at the centre. 
The energy density for both cases of uncharged and charged matter is a decreasing function so that $\rho^{\prime}<0$ within the star in Figure \ref{one}. In Figure \ref{two} the radial pressure $p_r$ is also a decreasing function; the pressure $p_r$ decreases slowly in the stellar core and then more rapidly in the stellar envelope. 
The radial pressure for the charged matter vanishes at the boundary so that $p_r(9.21)=0$. 
The tangential pressure $p_t$ represented in Figure \ref{three} is also a decreasing function. 
Similar profiles for $p_t$ have been presented in the treatments of \cite{sharma2007} and \cite{takisa2016}. 
In Figure \ref{four} we present the profile for the anisotropy $\Delta$. 
We notice that the anisotropy $\Delta$  vanishes at the centre, has a decreasing profile in the core, and then becomes an increasing function in the envelope.
Similar behaviour was obtained by \cite{takisa2016} in the absence of charge.
We represent the profile of the speed of sound $\frac{dp_r}{d\rho}$ in Figure \ref{five}.
It has the largest value in the core and the smallest value in the envelope. 
However for all values $\frac{dp_r}{d\rho}<1$ within the stellar object and the speed of sound is less than the speed of light.  In Figure \ref{six} the quantity $\frac{dp_t}{d\rho}$ remains  positive and finite. The quantity $\frac{dp_r}{d\rho}- \frac{dp_t}{d\rho}$ in Figure \ref{seven} is always positive and bounded by unity which is a necessary condition for stability. In Figure \ref{eight} we observe that the adiabatic index satisfies the condition $\Gamma > \frac{4}{3}$ which is the requirement  for a stable configuration. 
We notice that the presence of the electric field in all matter variables is greater in the envelope, from the matching interface to the stellar surface. This behaviour is consistent with the form of the electric field $E^2=\frac{sa^2 r^4}{(1+br^2)^2}$, which will be stronger in the spherical shell near the envelope surface.

\begin{table*}
\caption{ Mass-radius relationship of some pulsar stars. \label{table1-regular}}
\begin{tabular}{@{}cccccccccccccccc@{}}
%\begin{tabular}{lcccccccc }
\tableline
$\tilde{b}$&& $\alpha$ && $\gamma$&&$R_{I}$ &&$R_{II}$&&$M$&&$\frac{M}{R_{II}}$  & &STAR\\
&&  && && (km) &&(km) && $(M_\odot)$&& &&  \\ 
\tableline
-25.7351&&0.1029&&0.1245&&6.87&&10.30&&1.97 &&0.191&&PSR J1614-2230\\
(-25.7351)&&(0.1021)&&(0.0781)  && (6.87) && (9.21) && (2.13) &&(0.231) &&\\
-25.2941&&0.1044&&0.1237&&6.56 &&9.99&&1.77 &&0.177& &Vela X-1\\
(-25.2941)&&(0.1037)&&(0.0709)  && (6.56) && (8.96) && (1.92) && (0.214)&&\\
-25.055&&0.105&&0.1236 &&6.39&&9.82&&1.667 &&0.170&&PSR J1903+327\\
(-25.055)&&(0.1046)&&(0.0668)  && (6.39) && (8.82) && (1.81) && (0.205)&&\\
-24.6104&&0.1068&&0.1226&&6.08 &&9.51&&1.49 &&0.157&&Cen X-3\\
(-24.6104)&&(0.1062)&&(0.0593)  && (6.08) && (8.57) && (1.62) &&(0.189) &&\\
-24.023&&0.109&&0.1219&&5.70 &&9.13&&1.29 &&0.141&&SMC X-1\\
(-24.023)&&(0.1087)&&(0.0494)  && (5.70) && (8.25) && (1.40) &&(0.170) &&\\
\tableline
\end{tabular}
\end{table*}

\begin{table*}
\caption{Changing core radius values for PRS J1614-2230 structure. \label{table2-regular}}
\begin{tabular}{@{}cccccccccccccccccccccccc@{}}
%\begin{tabular}{lcccccccc }
\tableline
$\tilde{b}$ &&&&$\alpha$ &&&&$\gamma$ &&&&$ R_{I}$ &&&&$R_{II}$&&&&$M$  \\
&&&& &&&& &&&&    (km)&&&& (km)&&&& $(M_\odot)$\\ 
\tableline
-25.7351&&&&0.1029&&&&0.1245&&&&6.87&&&&10.30&&&&1.97\\
(-25.7351)&&&&(0.1021)&&&&(0.0781)&&&&(6.87)&&&&(9.21)&&&&(2.13)\\
-25.7351&&&&0.1029&&&&0.1381&&&&6.37&&&&10.30&&&&1.97\\
(-25.7351)&&&&(0.1021)&&&&(0.0918)&&&&(6.37)&&&&(9.21)&&&&(2.13)\\

-25.7351&&&&0.1029&&&&0.1490&&&&5.94&&&&10.30&&&&1.97 \\
(-25.7351)&&&&(0.1021)&&&&(0.1028)&&&&(5.94)&&&&(9.21)&&&&(2.13)\\
-25.7351&&&&0.1029&&&&0.1583&&&&5.54&&&&10.30&&&&1.97 \\
(-25.7351)&&&&(0.1021)&&&&(0.1123)&&&&(5.54)&&&&(9.21)&&&&(2.13)\\

-25.7351&&&&0.1029&&&&0.1670&&&&5.14&&&&10.30 &&&&1.97\\
(-25.7351)&&&&(0.1021)&&&&(0.1212)&&&&(5.14)&&&&(9.21)&&&&(2.13)\\
\tableline
\end{tabular}
\end{table*}

\begin{table*}
\caption{Changing envelope radius values for PRS J1614-2230 structure. \label{table3-regular}}
\begin{tabular}{@{}cccccccccccccccccccccccc@{}}
%\begin{tabular}{lcccccccc }
\tableline
$\tilde{b}$ &&&&$\alpha$ &&&&$\gamma$ &&&&$ R_{I}$ &&&&$R_{II}$&&&&$M$  \\
&&&& &&&& &&&&    (km)&&&& (km)&&&& $(M_\odot)$\\ 
\tableline
-25.7351&&&&0.1029&&&&0.1245&&&&6.87&&&&10.30&&&&1.97 \\
(-25.7351)&&&&(0.1021)&&&&(0.0781)&&&&(6.87)&&&&(9.21)&&&&(2.13)\\
-25.7351&&&&0.1188&&&&0.1943&&&&6.87&&&&10.80&&&&1.97 \\
(-25.7321)&&&&(0.1025)&&&&(0.0977)&&&&(6.87)&&&&(9.71)&&&&(2.13)\\
-25.7351&&&&0.1362&&&&0.2944&&&&6.87&&&&11.30&&&&1.97  \\
(-25.7351)&&&&(0.1028)&&&&(0.1185)&&&&(6.87)&&&&(10.21)&&&&(2.13)\\
-25.7351&&&&0.1554&&&&0.4348&&&&6.87&&&&11.80&&&&1.97  \\
(-25.7351)&&&&(0.1032)&&&&(0.1403)&&&&(6.87)&&&&(10.71)&&&&(2.13)\\
-25.7351&&&&0.1762&&&&0.6281&&&&6.87&&&&12.30 &&&&1.97 \\
(-25.7351)&&&&(0.1036)&&&&(0.1632)&&&&(6.87)&&&&(11.21)&&&&(2.13)\\
\tableline
\end{tabular}
\end{table*}

\begin{table*}
\caption{Gravitational redshifts for region $I$ with radii $R_I$ for the star PRS J1614-2230. \label{table4-regular}}
\begin{tabular}{@{}cccccccccccccccccccccccc@{}}
%\begin{tabular}{lcccccccc }
\tableline
$\tilde{b}$ &&&&$\alpha$ &&&&$\gamma$ &&&&$ R_{I}$ &&&&$Z_I$  \\
\tableline
-25.7351&&&&0.1029&&&&0.1245&&&&6.87&&&& 0.5312\\
(-25.7351)&&&&(0.1021)&&&&(0.0781)&&&&(6.87)&&&&(0.6224)\\
-25.7351&&&&0.1029&&&&0.1381&&&&6.37&&&& 0.6190\\
(-25.7351)&&&&(0.1021)&&&&(0.0918)&&&&(6.37)&&&&(0.7375)\\

-25.7351&&&&0.1029&&&&0.1490&&&&5.94&&&& 0.7234\\
(-25.7351)&&&&(0.1021)&&&&(0.1028)&&&&(5.94)&&&&( 0.8803)\\
-25.7351&&&&0.1029&&&&0.1583&&&&5.54&&&& 0.8608 \\
(-25.7351)&&&&(0.1021)&&&&(0.1123)&&&&(5.54)  &&&&(1.0804)\\

-25.7351&&&&0.1029&&&&0.1670&&&&5.14&&&&1.0696 \\
(-25.7351)&&&&(0.1021)&&&&(0.1212)&&&&(5.14)&&&&(1.4168)\\
\tableline
\end{tabular}
\end{table*}

\begin{table*}
\caption{Gravitational redshifts for region $II$ with radii $R_{II}$ for the star PRS J1614-2230. \label{table5-regular}}
\begin{tabular}{@{}cccccccccccccccccccccccc@{}}
%\begin{tabular}{lcccccccc }
\tableline
$\tilde{b}$ &&&&$\alpha$ &&&&$\gamma$ &&&&$R_{II}$&&&&$Z_{II}$  \\
\tableline
-25.7351&&&&0.1029&&&&0.1245&&&&10.30&&&&0.2725 \\
(-25.7351)&&&&(0.1021)&&&&(0.0781)&&&&(9.21)&&&&(0.3640)\\
-25.7351&&&&0.1188&&&&0.1943&&&&10.80&&&& 0.2547\\
(-25.7321)&&&&(0.1025)&&&&(0.0977)&&&&(9.71)&&&&(0.3348)\\
-25.7351&&&&0.1362&&&&0.2944&&&&11.30&&&&0.2391  \\
(-25.7351)&&&&(0.1028)&&&&(0.1185)&&&&(10.21)&&&&(0.3099)\\
-25.7351&&&&0.1554&&&&0.4348&&&&11.80&&&&0.2253  \\
(-25.7351)&&&&(0.1032)&&&&(0.1403)&&&&(10.71)&&&&(0.2886)\\
-25.7351&&&&0.1762&&&&0.6281&&&&12.30 &&&&0.2130 \\
(-25.7351)&&&&(0.1036)&&&&(0.1632)&&&&(11.21)&&&&(0.2700)\\
\tableline
\end{tabular}
\end{table*}

\section{\label{eee} Conclusion}

In this paper we used the core envelope setting in order to model a charged anisotropic compact object.
We have chosen a quadratic equation of state in the stellar core and a linear equation of state in the stellar envelope. 
The radial pressure in the core is higher than the radial pressure in the envelope. 
We proved that the core region, the envelope region and the exterior spacetime match smoothly at their respective interfaces. Different masses, radii and compactification factors of the five stars PSR J1614-2230, PSR J1903+0327, Vela X-1, SMC X-1, Cen X-3 were obtained in the presence of electric field. These results are in agreement with the treatments of \cite{Mafa2014a}, \cite{Mafa2014b}, \cite{takisa2016} and \cite{azam2015}. We plotted the matter variables related to the compact object PSR J1614-2230. The matter variables and the metric potentials are well behaved within the star, and there is smooth matching between the core and the envelope. We demonstrated that values for the radius of the core and the radius of the envelope can change; this was achieved by choosing different parameter values. This made it possible for us to take in account the core envelope models with varying compactification factors in the stellar core and the stellar envelope. The main result of our study is that it possible to have a charged anisotropic regular core envelope relativistic model with equations of state in both the core region and the envelope region. In future investigation it would be interesting to investigate the effect of other forms of equations of state and different choices of the electric field.
\newpage
\begin{center}
\textbf{Acknowledgements }
\end{center}
PMT thanks the National Research Foundation and the University of
KwaZulu-Natal for financial support.
SDM acknowledges that this work is based upon research supported by the South African Research
Chair Initiative of the Department of Science and
Technology and the National Research Foundation.

\end{document}